\magnification = 1120
\baselineskip = 15 pt

\def \noi {\noindent}

\def \P {{\bf P}}
\def \E {{\bf E}}

\def \al {\alpha}

\def \sig {\sigma}
\def \del {\delta}
\def \kap {\kappa}
\def \eps {\varepsilon}

\def \ws {w_s}

\def \s {{\bf s}}
\def \b {{\bf b}}

\def \R {{\bf R}}
\def \D {{\bf D}}
\def \phio {\phi^{(0)}}
\def \phil {\phi^{(1)}}

\def\sqr#1#2{{\vcenter{\vbox{\hrule height.#2pt\hbox{\vrule width.#2pt height#1pt \kern#1pt\vrule width.#2pt}\hrule height.#2pt}}}}

\def \square{\hfill\mathchoice\sqr56\sqr56\sqr{4.1}5\sqr{3.5}5}

\def \qed {$\square$ \medskip}

\def \sect#1{\bigskip \noindent {\bf  #1} \medskip}

\def \th#1#2{\medskip \noindent {\bf  Theorem #1.}   \it #2 \rm \medskip}
\def \prop#1#2{\medskip \noindent {\bf  Proposition #1.}   \it #2 \rm \medskip}
\def \cor#1#2{\medskip \noindent {\bf  Corollary #1.}   \it #2 \rm \medskip}
\def \pf {\noindent  {\it Proof}.\quad }
\def \lem#1#2{\medskip \noindent {\bf  Lemma #1.}   \it #2 \rm \medskip}
\def \ex#1{\medskip \noindent {\bf  Example #1.}}
\def \rem#1 {\medskip \noi {\bf Remark #1.  }}
\def \as#1 {\medskip \noi {\bf Assumption #1.  }}

\centerline{\bf  Optimal Investment to Minimize the Probability of Drawdown} \bigskip \bigskip

\noi Bahman Angoshtari

Email: bango@umich.edu

\medskip

\noi Erhan Bayraktar

Email: erhan@umich.edu

\medskip

\noi Virginia R. Young

Email: vryoung@umich.edu

\bigskip

\noi 530 Church Street

\noi Department of Mathematics, University of Michigan

\noi Ann Arbor, Michigan, 48109

\bigskip

\centerline{Version: 11 December 2015}

\bigskip

\noindent{\bf  Abstract:}  We determine the optimal investment strategy in a Black-Scholes financial market to minimize the so-called {\it probability of drawdown}, namely, the probability that the value of an investment portfolio reaches some fixed proportion of its maximum value to date.  We assume that the portfolio is subject to a payout that is a deterministic function of its value, as might be the case for an endowment fund paying at a specified rate, for example, at a constant rate or at a rate that is proportional to the fund's value.

\medskip


\sect{1. Introduction}

We determine the optimal investment strategy in a Black-Scholes financial market to minimize the so-called {\it probability of drawdown}, namely, the probability that the value of the portfolio reaches some fixed proportion of its maximum value to date.  We assume that the portfolio is subject to a payout that is a deterministic function of the value of the fund, as might be the case for an endowment fund paying at a specified rate.

Our paper falls naturally within the area of optimally controlling wealth to reach a goal.  Research on this topic began with the seminal work of Dubins and Savage (1965, 1976) and continued with the work of Pestien and Sudderth (1985), Orey et al. (1987), Sudderth and Weerasinghe (1989), Kulldorff (1993), Karatzas (1997), Browne (1997, 1999a, 1999b), Young (2004), and Bayraktar and Young (2015).  A typical problem considered in this research is to control a process to maximize the probability the process reaches $b$, either before a fixed time $T$, such as in Karatzas (1997), or before the process reaches $a < b$, such as in Pestien and Sudderth (1985).  Pestien and Sudderth show that to maximize the probability of reaching $b$ before $a$, one maximizes the ratio of the drift of the diffusion divided by its volatility squared.

B\"auerle and Bayraktar (2014) prove, via pathwise arguments, that maximizing the ratio of the drift of a diffusion divided by its volatility squared minimizes the probability of ruin, as in Pestien and Sudderth (1985).  The additional pathwise arguments demonstrate, among other things, that the same optimizer also optimizes any decreasing measurable function of the running minimum of a given, but arbitrary, diffusion. The latter was earlier observed in Bayraktar and Young (2007) when the state variable was the wealth process in a Black-Scholes investment model with an arbitrary consumption function.  B\"auerle and Bayraktar (2014) also showed that if this maximized ratio is independent of the state of the process, then the probability of drawdown is also a minimum.

In this paper, we work in the set up of  Bayraktar and Young (2007), in which this assumption of independence is not satisfied. We use a verification argument to prove that the investment strategy that minimizes the probability of ruin also minimizes the probability of drawdown.  We did not expect this result because when minimizing the probability of drawdown, the ``ruin level'' is a non-decreasing process.  However, it is not obvious that if the ruin level changes over time, as it does when minimizing the probability of drawdown, then the investment strategy to minimize the probability of ruin will also minimize the probability of drawdown, but that result is exactly what we show in this paper.

In other work involving drawdown, such as Elie and Touzi (2008), drawdown has been used as a constraint in maximizing expected utility.  By contrast, we minimize the probability of drawdown directly, recognizing that the manager of an investment or endowment fund might first choose a payout function and then seek to manage the fund so that the value of the fund does not fall below some given proportion of its maximum.

The rest of the paper is organized as follows. In Section 2, we introduce the financial market, define the problem of minimizing the probability of drawdown, and compute the minimum probability of drawdown, $\phi$, when the maximum portfolio value is greater than the so-called {\it safe level}.  In Section 3, we prove a verification theorem for the minimum probability of drawdown, $\phi$, and use it to compute $\phi$ when the maximum portfolio value is less than the safe level.  We learn that the optimal investment strategy for minimizing the probability of drawdown is {\it identical} to the strategy for minimizing the probability of ruin, for any ruin level.  In fact, the same investment strategy will minimize the expectation of any function that is non-increasing with respect to the minimum portfolio value and non-decreasing with respect to the maximum portfolio value.  Finally, in Section 4, we examine the behavior of the optimally controlled portfolio value.

\sect{2. Problem statement and preliminary result}

In this section, we first present the financial ingredients that make up the value of the investment or endowment fund, namely, a payout rate, a riskless asset, and a risky asset.  We, then, define the minimum probability of drawdown and compute that probability explicitly when the maximum portfolio value is greater than the safe level.

We assume that the manager of the fund invests in a riskless asset that earns interest at a constant rate $r > 0$.  Also, the manager invests in a risky asset whose price at time $t$, $S_t$, follows geometric Brownian motion given by
$$
\left\{
\eqalign{
dS_t &= \mu S_t dt + \sig S_t dB_t, \cr
S_0 &= S > 0,} \right.
$$
in which $\mu > r$, $\sig > 0$, and $B$ is a standard Brownian motion with respect to a filtration of a probability space $(\Omega, {\cal F}, \P)$.  Let $W_t$ be the value at time $t$ of the investment fund, and let $\pi_t$ be the amount that the manager invests in the risky asset at that time.  It follows that the amount invested in the riskless asset is $W_t - \pi_t$.  We assume that the fund pays out at a deterministic rate $c(W_t) \ge 0$.\footnote{$^1$}{We are thinking of this payout as money given to recipients of an endowment fund or to shareholders of an investment fund.}    Therefore, the portfolio value follows the process
$$
dW_t = \left[rW_t + (\mu - r) \pi_t  - c(W_t) \right] dt + \sig \pi_t dB_t,
\eqno(2.1)
$$
with $W_0 = w$ given.

Define the maximum portfolio value $M_t$ at time $t$ by
$$
M_t = \max \left[ \sup_{0 \le s \le t} W_s, \; M_0 \right],
$$
\noindent in which we include $M_0 = m > 0$ (possibly different from $W_0 = w$) to allow the investment fund to have a financial past.  By {\it drawdown}, we mean that the value of the fund reaches $\al \in (0, 1)$ times its maximum value.  Define the corresponding hitting time by $\tau_\al := \inf\{t \ge 0: W_t \le \al M_t\}$.  As an aside, if $\al = 0$, then we are in the case of minimizing the probability of ruin for the fixed ruin level $0$, which has been considered in previous work; see Pestien and Sudderth (1985) and more recently Bayraktar and B\"auerle (2014).  Thus, without loss of generality, we require $\al \in (0, 1)$.\footnote{$^2$}{Angoshtari et al.\ (2015) consider a similar problem, but they minimize the expected time spent in drawdown for a finitely-lived agent when $c(w) = \kappa w$.}

When the payout rate is constant (which is the case for many endowment funds), say, identically $c$ per year, then the fund level $c/r$ plays a special role.  If the value of the investment fund is at least equal to $c/r$, then the fund manager can invest all wealth in the riskless asset and earn at least $c$ per year, and the fund's value will never decrease.  In particular, drawdown cannot occur in this case.  We generalize from this special case in the following assumption.

\as{2.1} {Throughout this paper, we assume that the payout function $c(w)$ is a continuous, non-negative, non-decreasing function of the value of the endowment fund on $(0, \infty)$, such that there exists a unique $\ws \in (0, \infty]$ for which\footnote{$^3$}{We assume uniqueness of $\ws$ for mathematical simplicity.}
$$
rw < c(w), \quad \hbox{ for all } w < \ws,
$$
and
$$
rw > c(w), \quad \hbox{ for all } w > \ws.
$$
We allow $\ws = \infty$, which is the case, for example, if $c(w) = \kap w$ with $\kap > r$.  If $W_0 = w \ge \ws$, then we can set $\pi_t = 0$ for all $t \ge 0$, which implies
$$
dW_t = (rW_t - c(W_t)) \, dt \ge 0.
$$
Under this investment strategy, the value of the investment fund is non-decreasing, so drawdown cannot occur.  For this reason, we call $\ws$ the {\it safe level}.    \qed}

Denote the minimum probability of drawdown by $\phi(w, m)$, in which the arguments $w$ and $m$ indicate that one conditions on the current value $w$ of the investment fund, with maximum (past) value $m$.  Specifically, $\phi$ is the minimum probability that $\tau_\al < \infty$, in which one minimizes with respect to admissible investment strategies $\pi$.  A strategy $\pi$ is {\it admissible} if it is ${\cal F}_t$-progressively measurable (in which ${\cal F}_t$ is the augmentation of $\sig(W_s: 0 \le s \le t)$) and if it satisfies the integrability condition $\int_0^t \pi_s^2 \, ds < \infty$ almost surely, for all $t \ge 0$.  Thus, $\phi$ is formally defined by
$$
\phi(w, m) = \inf_{\pi} \P^{w, m} \left(\tau_\al < \infty \right) = \inf_{\pi} \E^{w, m} \left( {\bf 1}_{\{\tau_\al < \infty\}} \right),
\eqno(2.2)
$$
for $w \le m$.  Here, $\P^{w, m}$ and $\E^{w, m}$ denote the probability and expectation, respectively, conditional on $W_0 = w$ and $M_0 = m$.

We first consider the case for which $m \ge \ws$, which implies that $\ws < \infty$.  If $W_0 = w \ge \ws$, then drawdown is impossible; thus, we assume that $W_0 = w \in (\al m, \ws)$.  $W_0 = w < \ws$ implies that either $W_t < \ws$ almost surely, for all $t \ge 0$, or $W_t = \ws$ for some $t > 0$.\footnote{$^4$}{If the latter occurs, set $\pi_s = 0$ for all $s \ge t$, as discussed in Assumption 2.1.}  In either event, $M_t = m$ almost surely, for all $t \ge 0$, and avoiding drawdown is equivalent to avoiding ruin with a (fixed) ruin level of $\al m$.  B\"auerle and Bayraktar (2014) show that the optimal investment strategy in this case is the one that maximizes the ratio of the drift of the value process in (2.1) to its volatility squared.  Thus, when the value equals $W_t = w$, the optimal amount to invest in the risky asset is given in feedback form by
$$
\pi^*(w) = {2 \left( c(w) -  r w \right) \over \mu - r},
\eqno(2.3)
$$
independent of $m$ and $\al$.  One can show, under the investment strategy given in (2.3), the value of the investment account follows the process
$$
dW_t = \left( c(W_t) - r W_t \right) \left\{  dt + {2 \sig \over \mu - r} \, dB_t \right\}.
\eqno(2.4)
$$

From B\"auerle and Bayraktar (2014, Theorem 4.1), we know that the probability of ruin under this investment strategy is given by
$$
h(w, m) = 1 - {g(w, m)  \over  g(\ws, m)} \; ,
\eqno(2.5)
$$
in which $g$ is defined by
$$
g(w, m) = \int_{\al m}^w \exp \left( - \int_{\al m}^y {\del \, du \over c(u) - ru} \right) \, dy,
\eqno(2.6)
$$
with
$$
\del = {1 \over 2} \left( {\mu - r \over \sig} \right)^2.
$$
Note that $g$ in (2.6) is the {\it scale function} associated with the diffusion in (2.4); see Karatzas and Shreve (1991, page 339).  Also, note that because $\ws$ is necessarily finite in this section, $g(\ws, m)$ is also finite.  Indeed, the integrand in the expression for $g$ is bounded above by $1$; thus, $g(\ws, m) \le \ws - \al m < \infty$.

In the next proposition, we summarize the above discussion.

\prop{2.1} {The minimum probability of drawdown $\phi$ on $\{ (w, m) \in (\R^+)^2: \al m \le w \le \ws \le m \}$ is given by the expression in $(2.5)$.  The optimal amount to invest in the risky asset when $W_t = w$ is given by $(2.3)$, independent of $m$ and $\al$.  \qed}

It follows from the investment strategy given in (2.3), that as the portfolio value increases towards $\ws$, the amount invested in the risky asset approaches zero. This makes sense because as the value of the investment account increases, the manager does not need to take on as much risk to achieve the limiting payout rate of $c(\ws)$.

\sect{3.  Minimum probability of drawdown when the maximum portfolio value $m < \ws$}

In the previous section, we showed that it is optimal for $M_t = m$ almost surely, for all $t \ge 0$, when $\al m < w < \ws \le m$.  In this section, we show that allowing $M$ to increase above $m$ is optimal when $m < \ws$.  To prove this result, we rely on a verification lemma.  First, define the differential operator ${\cal L}^\beta$ for $\beta \in \R$ by
$$
{\cal L}^\beta f = (rw + (\mu - r) \beta - c(w)) f_w + {1 \over 2} \sig^2 \beta^2 f_{ww},
$$
in which $f = f(w, m)$ is twice-differentiable with respect to its first variable.

Assume that $w > \al m$; otherwise, drawdown has occurred, and the game is over.  Thus, in general, we need only consider $\phi$ on the domain $\D := \{(w,m) \in ({\bf R}^+)^2: \al m \le w \le \min(m, \ws) \}$, in which we allow $\ws = \infty$.

\lem{3.1} {Suppose $h: {\D} \rightarrow {\bf R}$ is a bounded, continuous function that satisfies the following conditions:
\item{$(i)$} $h(\cdot, m) \in C^2((\al m, \min(m, \ws)))$ is a non-increasing, convex function,
\item{$(ii)$} $h(w, \cdot)$ is continuously differentiable, except possibly at $\ws$ when $\ws < \infty$, where it has right- and left-derivatives,
\item{$(iii)$} $h_m(m, m) \ge 0$ if $m < \ws$,
\item{$(iv)$} $h(\al m, m) = 1$,
\item{$(v)$} $h(\ws, m) = 0$ if $m \ge \ws$ $($in the limiting sense if $\ws = \infty)$,
\item{$(vi)$} ${\cal L}^\beta h \ge 0$ for all $\beta \in {\bf R}$.}

\noi{\it Then, $h(w, m) \le \phi(w, m)$ on $\D$.}

\medskip

\pf Assume that $h$ satisfies the conditions specified in the statement of this theorem.  Let $W^\pi$ and $M^\pi$ denote the portfolio value and the maximum portfolio value, respectively, when the manager uses an admissible investment policy $\pi$.  Also, assume that the ordered pair of initial value and maximum value $(w, m)$ lie in $\D$.

Fix an admissible investment policy $\pi$.  Define $\tau_n = \inf \{t \ge 0:  \int_0^t \pi^2_s \, ds \ge n \}$, $\tau_{\ws} = \inf \{ t \ge 0: W_t^\pi = \ws \}$, and $\tau = \tau_\al \wedge \tau_n \wedge \tau_{\ws}$.  By applying It\^{o}'s formula to $h(w, m)$, we have
$$
\eqalign{
h(W^\pi_{\tau}, M^\pi_{\tau}) &= h(w, m) + \int_0^{\tau} h_w(W^\pi_t, M^\pi_t) \, \sig \, \pi_t \, dB_t  \cr
& \quad +\int_0^{\tau} {\cal L}^\pi h(W^\pi_t, M^\pi_t) \, dt  + \int_0^{\tau} h_m (W^\pi_t, M^\pi_t) \, dM^\pi_t.}
\eqno(3.1)
$$

It follows from the definition of $\tau_n$ that
$$
\E^{w, m} \left[\int_0^{\tau} h_w(W^\pi_t, M^\pi_t) \, \sig \, \pi_t \, dB_t \right] = 0.
$$
Also, the second integral in (3.1) is non-negative because of condition (vi) of the theorem.  Finally, the third integral  is non-negative almost surely because $dM_t$ is non-zero only when $M_t = W_t$ and $h_m(m,m) \ge 0$.   Here, we also used the fact that $M$ is non-decreasing; therefore, the first variation process associated with it is finite almost surely, and we conclude that the cross variation of $M$ and $W$ is zero almost surely.  Thus, we have
$$
\E^{w,m} \left[h(W^\pi_{\tau}, M^\pi_{\tau}) \right] \ge h(w, m).
$$

Because $h$ is bounded by assumption, it follows from the dominated convergence theorem that
$$
\E^{w,m}\left[h \left(W^\pi_{\tau_\al \wedge \tau_{\ws}}, M^\pi_{\tau_\al \wedge \tau_{\ws}} \right) \right] \ge h(w,m).
$$
Since $W^\pi_{\tau_\al} = \al M^\pi_{\tau_\al}$ and $W^\pi_{\tau_{\ws}} = \ws$ when $(W^\pi_0, M^\pi_0) = (w, m) \in {\D}$, it follows from conditions (iv) and (v) of the theorem that
$$
h(w, m) \le \E^{w,m} \left( {\bf 1}_{\{\tau_\al < \tau_{\ws}\}} \right) = \E^{w,m} \left( {\bf 1}_{\{\tau_\al < \infty \}} \right).
\eqno(3.2)
$$
The equality in (3.2) follows from the fact that $\tau_\al = \infty$ if $\tau_{\ws} \le \tau_\al$.  By taking the infimum over admissible investment strategies, and by applying the second representation of $\phi$ from (2.2), we obtain $h \le \phi$ on $\D$.  \qed

\cor{3.2}{Suppose $h$ satisfies the conditions of Lemma $3.1$ in such a way that condition $(iii)$ holds with equality and that $(vi)$ holds with equality for some admissible strategy $\pi$ defined in feedback form by $\pi_t = \pi(W_t, M_t),$ in which we slightly abuse notation.  Specifically, suppose ${\cal L}^{\pi(w, m)} h(w, m) = 0$ on $\D$.  Then, $h(w, m) = \phi(w, m)$ on $\D$, and $\pi$ is an optimal investment strategy.}

\pf In the proof of Lemma 3.1, if we have equality in conditions (iii) and (vi), then we can conclude that $h = \phi$ on $\D$.  \qed

Consider the following boundary-value problem, the limit of whose solution, according to Corollary 3.2, is a candidate for the minimum probability of drawdown.  Let $N \le \ws$; then, on $\al m \le w \le m \le N$,
$$
\left\{
\eqalign{
&(rw - c(w)) h^N_w - \del {(h^N_w)^2 \over h^N_{ww}} = 0, \cr
&h^N(\al m, m) = 1, \quad h^N_m(m, m) = 0, \cr
&h^N(N, N) = 0.}
\right.
\eqno(3.3)
$$
In the next proposition, we present the solution of (3.3).

\prop{3.3} {The solution of $(3.3)$ on $\{ (w, m) \in (\R^+)^2: \al m \le w \le m \le N \}$ is given by
$$
h^N(w, m) = 1 -   e^{- \int_m^{N} f(y) dy} \; {g(w, m) \over g(N, N)},
\eqno(3.4)
$$
in which $f$ is defined by
$$
f(m) = \al \left[  {1 \over g(m, m)} -  {\del \over c(\al m) - r \al m} \right],
\eqno(3.5)
$$
and $g$ is given in $(2.6)$.}

\pf   It is straightforward to show that $h$ in (3.4) satisfies the differential equation in (3.3), as well as the boundary conditions $h^N(\al m, m) = 1$ and $h^N(N, N) = 0$.  Also, $h^N_m(m, m) = 0$; indeed,
$$
\eqalign{
h^N_m(w, m) &= - \, e^{- \int_m^{N} f(y) dy} \; {f(m) g(w, m) + g_m(w, m) \over g(N, N)} \cr
&\propto -f(m) g(w, m) + \al \left[ 1  + {\del \over r \al m - c(\al m)} \, g(w, m) \right] = \al \left[ 1 - {g(w, m) \over g(m, m)} \right], 
}
$$
which equals $0$ when $w = m$. \qed

\rem{3.1} {If $\ws < \infty$, then $h^{\ws}$ equals the minimum probability of drawdown on $\al m \le w \le m \le \ws$.  If $\ws = \infty$, then
$$
\lim_{N \to \infty} {e^{- \int_m^{N} f(y) dy} \over g(N, N)}
$$
might be indeterminate, so we have to take care with the limit. \qed}

We have the following theorem that combines the results of Propositions 2.1 and 3.3.

\th{3.1} {Let $g(\ws, \ws)$ denote $\lim_{m \to \ws-} g(m, m)$.  The minimum probability of drawdown $\phi$ on ${\D} := \{(w,m) \in ({\bf R}^+)^2: \al m \le w \le \min(m, \ws) \}$ equals
$$
\phi(w, m) = \cases{
1 - {g(w, m)  \over  g(\ws, m)}, &if $\al m \le w \le \ws, m \ge \ws$, \cr \cr
1 -  k(m) \, {g(w, m) \over g(\ws, \ws)}, &if $\al m \le w \le m < \ws$, \cr \cr
}
\eqno(3.6)
$$
in which $g$ is given by $(2.6)$, and in which $k$ is defined by
$$
k(m) = e^{- \int_m^{\ws} f(y) dy},
\eqno(3.7)
$$
with $f$ given by $(3.5)$.  The optimal amount to invest in the risky asset when $W_t = w$ is given by
$$
\pi^*(w) = {2 \left( c(w) -  r w \right) \over \mu - r},
\eqno(3.8)
$$
independent of $m$ and $\al$.}

\pf  Given Corollary 3.2 and the results of Propositions 2.1 and 3.3, the only item remaining to show is that the expression given in (3.6) is continuous at $m = \ws$ if $\ws < \infty$, which is clear.  \qed

\rem{3.2} {Even though the maximum portfolio value is allowed to increase, the optimal investment strategy for minimizing the probability of drawdown when $m < \ws$ is {\it identical} to the optimal investment strategy when $m \ge \ws$, and both equal the optimal investment strategy to minimize the probability of ruin.  In fact, the same investment strategy will minimize the expectation of any function that is non-increasing with respect to the minimum portfolio value and non-decreasing with respect to the maximum portfolio value.  Indeed, the differential equation would remain the same; the only change would come in the various boundary conditions at the minimum and maximum.   \qed}

\rem{3.3} {It is not obvious {\it a priori} that the optimal strategy will be such that the individual seeks to reach the upper bound, $w_s$, before the (moving) lower bound $\alpha m$.  In fact, when the horizon is not infinite, it can be optimal to invest in such a way that $M_t$ does not increase above its current maximum. For example, see the working paper of Angoshtari et al.\ (2016).  In the setting of that paper, when an individual seeks to minimize the probability of drawdown before death, if maximum wealth is small enough, then it is optimal {\it not} to allow maximum wealth to increase; see Theorem 5.5.  Thus, when the horizon is finite (with probability one), it might be optimal to keep maximum wealth constant and to gamble on dying before drawdown occurs.  However, when the horizon is infinite (as in this paper), then Theorem 3.1 shows that it is optimal to allow maximum wealth to increase to the safe level. \qed}

\sect{4. Optimally controlled portfolio value}

We end our paper by considering the behavior of the optimally controlled portfolio value.  Because the optimal investment strategy is independent of the maximum $m$, we can effectively ignore the value of $m$.

First, consider the case for which $\ws < \infty$, regardless of the value of $m$.  Extend $c$ to all of $\R$ by setting $c(w) = c(0)$ for $w < 0$, and extend $\pi^*$ given in (2.3) to all of $\R$ as follows:
$$
\pi^*(w) = 
\cases{ {c(w) - rw \over \mu - r}, &if $w < \al m$, \cr \cr
{2 \left( c(w) -  rw \right) \over \mu - r}, &if $\al m \le w \le \ws$, \cr \cr
0, &if $w > \ws$.}
\eqno(4.1)
$$
Then, define $\s(w) = \sig \pi^*(w)$ and $\b(w) = rw + (\mu - r) \pi^*(w) - c(w)$; note that $\b(w) = 0$ for $w < \al m$.  Define $p$ on $\R$ by
$$
p(w) = \int_{\al m}^w \exp \left( -2 \int_{\al m}^y {\b(z) \, dz \over \s^2(z)} \right) \; dy.
\eqno(4.2)
$$
Note that $p(w) = g(w, m)$ for $\al m \le w \le \ws$ and $p(-\infty) = \int_{\al m}^{-\infty} 1 \, dy = -\infty$.  Also, define $v$ on $\R \times \R^+$ by
$$
\eqalign{
v(w, m) &= \int_{\al m}^w p'(y) \int_{\al m}^y {2 \, dz \over p'(z) \, \s^2(z)} \; dy = \int_{\al m}^w \int_{\al m}^y {2 \over \s^2(z)} \, \exp \left( - 2 \int_{z}^y {\b(u) \, du \over \s^2(u)} \right) \, dz \, dy.}
\eqno(4.3)
$$
Compare the expression for $v$ with the one in (5.65) of Karatzas and Shreve (1991, page 347).  Because $p(-\infty) = -\infty$, it follows that $v(-\infty, m) = \infty$.

Because $c(w) > rw$ for $w < \ws$ and $c(\ws) = r \ws$ when $\ws$ is finite, we generally expect $c(w) - rw$ to decrease to $0$ in a left-neighborhood of $\ws$ for most reasonable payout functions.  In the following proposition, we show that if the payout function has a bounded (negative) derivative near $\ws$, then $v(\ws^-, m) = \infty$.

\prop{4.1} {Suppose $\ws < \infty$.  In addition to Assumption $2.1$, assume that the payout function $c$ is continuously differentiable in a left-neighborhood of $\ws$ and that there exist constants $\eps \in (0, \ws - \al m)$ and $K > 0$ such that
$$
-K < c^\prime(w) - r < - {1 \over K}, \quad \forall w \in (\ws - \eps, \ws).
\eqno(4.4)
$$
Then, $v(\ws^-, m) = \infty$.}

\pf  Assume otherwise; specifically, assume that $v(\ws^-,m) < \infty$. For all $w \in (\ws - \eps, \ws)$, Fubini's theorem yields
$$
\eqalign{
v(w,m) &= \int_{\al m}^w \int_{\al m}^y {\del \over (c(z) - rz)^2} \exp \left(- \int_z^y {\del \, du \over c(u) - ru} \right) dz \, dy \cr \cr
&= \int_{\al m}^w {\del \over (c(z) - rz)^2} \int_z^w \exp \left(- \int_z^y {\del \, du \over c(u) - ru} \right) dy \, dz \cr \cr
&\ge \int_{\ws - \eps}^w {\del \over (c(z) - rz)^2} \int_z^w \exp \left(- \int_z^y {\del \, du \over c(u) - ru} \right) dy \, dz.
}
$$
Next, we find a lower bound for the inner integral. By (4.4),
$$
{-\del \over c(u) - ru} \ge \del K \, {c^\prime(u) - r \over c(u) - ru}, \quad \forall u \in (\ws - \eps, \ws).
$$
It, then, follows that
$$
\eqalign{
v(w,m) &\ge \int_{\ws-\eps}^w {\del \over (c(z) - rz)^2} \int_z^w \exp \left( \del K\int_z^y {c^\prime(u)-r \over c(u)-r u} \, du \right) dy \, dz \cr\cr
&= \del \int_{\ws-\eps}^w \left(c(z)-r z\right)^{-\del K - 2} \int_z^w \left(c(y)-r y\right)^{\del K} dy \, dz\cr\cr
&\ge - {\del \over K} \int_{\ws-\eps}^w \left(c(z)-r z\right)^{-\del K - 2} \int_z^w \left(c(y)-r y\right)^{\del K} (c^\prime(y) - r) \, dy \, dz\cr\cr
&= {\del \over K(\del K + 1)} \int_{\ws-\eps}^w \left(c(z)-r z\right)^{-\del K - 2} \Big(\left(c(z)-r z\right)^{\del K + 1} - \left(c(w)-r w\right)^{\del K + 1} \Big) \, dz\cr\cr
&= {\del \over K(\del K + 1)} \bigg(\int_{\ws-\eps}^w {dz \over c(z)-r z} - \left(c(w)-r w\right)^{\del K + 1} \int_{\ws-\eps}^w \left(c(z)-r z\right)^{-\del K - 2} dz \bigg)\cr\cr
&\ge {\del \over K(\del K + 1)} \bigg({1 \over r}\int_{\ws-\eps}^w {1 \over \ws - z} dz \cr\cr
&\hphantom{\ge {\del \over K(\del K - 1)} \bigg(}
+ K \left(c(w)-r w\right)^{\del K + 1} \int_{\ws-\eps}^w \left(c(z)-r z\right)^{-\del K - 2}(c^\prime(z) - r) \, dz \bigg) \cr\cr
&= {\del \over rK(\del K + 1)}\ln\left({\eps \over \ws-w}\right) + {\del \over (\del K + 1)^2} \bigg(\left({c(w) - r w \over c(\ws-\eps) - r (\ws-\eps)}\right)^{\del K + 1} - 1 \bigg).
}
$$
\smallskip

\noi For the second inequality above, we used (4.4), and for the third inequality, we used (4.4) and $0<c(z) - r z \le r(\ws-z)$ for $z<\ws$, which follows from Assumption 2.1. In particular, we have found a lower bound for $v(w, m)$ that approaches $\infty$ as $w \to \ws^-$. This contradicts our initial assumption of $v(\ws^-,m) < \infty$; thus, $v(\ws^-, m) = \infty$. \qed

\rem{4.1} {Let $S$ denote the first hitting time of $\al m$ and $\ws$ when the initial portfolio value $w$ lies in $(\al m, \ws)$.  From Proposition 5.5.32 of Karaztas and Shreve (1991, page 350), we deduce that if $v(\ws^-, m) = \infty$, then $0 < \P^w (S < \infty) < 1$.  Furthermore, if $v(\ws^-, m) = \infty$, then, because $v(-\infty, m) = \infty$, it follows from Feller's test for explosions (Theorem 5.5.29 of Karatzas and Shreve (1991, page 348)) that the optimally controlled portfolio value {\it never} reaches the safe level $\ws$.

Optimally controlled wealth follows the process given in (2.4).  As wealth approaches $w_s$, then both the drift and the volatility approach $0$.  Thus, it is reasonable to expect that the safe level might not be reachable, and the conditions of Proposition 4.1 give an instance when $w_s$ is not reachable.

In combination with $0 < \P^w (S < \infty) < 1$, it follows that either drawdown occurs (with probability $\phi(w, m) = \P^w (S < \infty)$) or the optimally controlled portfolio value lies strictly between $\al m$ and $c/r$, almost surely, for all time (with probability $1 - \phi(w, m)$).  \qed}

We present the following example that demonstrates that we can have $v(\ws^-, m) = \infty$ when (4.4) does not hold.

\ex{4.1} {Suppose $c(w) = r w + b(\ws - w)^2$ for some $b > 0$ and $\ws > 0$.  Note that $c'(w) = r - 2b(\ws - w)$; thus, to ensure that $c'(w) \ge 0$, assume that $\ws - {r \over 2b} \le \al m \le w \le \ws$.  (For example, one can set $b = {r \over 2 \ws}$ so that $\ws - {r \over 2b} = 0$.)  Then,
$$
g(w, m) = e^{{\del \over b(\ws - \al m)}} \int_{\al m}^w e^{- {\del \over b(\ws - y)}} \, dy,
$$
and
$$
\eqalign{
v(w, m) &= \left( {1 \over b} \left( {1 \over \ws - w} - {1 \over \ws - \al m} \right) + {2 \over \del} \, \ln \left( {\ws - w \over \ws - \al m} \right) + {2b \over \del^2} \, (w - \al m) \right) \cr
&\qquad - e^{{\del \over b(\ws - \al m)}} \left( {1 \over b} \, {1 \over (\ws - \al m)^2} - {2 \over \del} \, {1 \over \ws - \al m} + {2b \over \del^2} \right) \, \int_{\al m}^w e^{- {\del \over b(\ws - y)}} \, dy \cr \cr
&\ge \left( {1 \over b} \left( {1 \over \ws - w} - {1 \over \ws - \al m} \right) + {2 \over \del} \, \ln \left( {\ws - w \over \ws - \al m} \right) + {2b \over \del^2} \, (w - \al m) \right) \cr
&\qquad - e^{{\del \over b(\ws - \al m)}} \left( {1 \over b} \, {1 \over (\ws - \al m)^2} - {2 \over \del} \, {1 \over \ws - \al m} + {2b \over \del^2} \right) \, (w - \al m),
}
$$
\smallskip

\noi which implies that $v(\ws, m) = \infty$.  Thus, as discussed in Remark 4.1, the optimally controlled portfolio value never reaches the safe level $\ws$. \qed}

\rem{4.2} {If the payout function $c$ has a Taylor series expansion about $w = \ws$, then we can write
$$
c(w) = c(\ws) + (w - \ws) c'(\ws) + {1 \over 2} (w - \ws)^2 c''(\ws) + {\cal O} \left((w - \ws)^3 \right),
$$
or equivalently,
$$
c(w) - rw = - (\ws - w) (c'(\ws) - r) + {1 \over 2} (\ws - w)^2 c''(\ws) +  {\cal O} \left((w - \ws)^3 \right),
$$
for $w$ in a left-neighborhood of $\ws$.  Proposition 4.1 shows that, when $0 < | c'(\ws) - r | < \infty$, the safe level is never reached.  Example 4.1 shows that, if $c'(\ws) = r$, it can still be true that the safe level is never reached.  In general, we expect that the safe level is never reached, that is, $v(\ws^-, m) = \infty$, and we welcome the interested reader to prove this statement.  \qed}

Next, suppose $\ws = \infty$.  In the following proposition, we show that if $c(w) - rw$ is eventually bounded away from $0$ as the portfolio value increases, then the probability of drawdown is identically $1$.  First, we prove a comparison result that we will use in the proof of the proposition.

\lem{4.2} {Suppose the payout functions are ordered, specifically, suppose $c_0 \le c_1$.  Denote the corresponding minimum probabilities of drawdown by $\phio$ and $\phil$.  Then, $\phio \le \phil$ on $\D^{(0)} := \left\{(w,m) \in ({\bf R}^+)^2: \al m \le w \le \min \left(m, \ws^{(0)} \right) \right\}$, in which $\ws^{(0)} \le \ws^{(1)}$ are the values of $\ws$ that correspond to $c_0$ and $c_1$, respectively.}

\pf Define the function $F$ by
$$
F(w, f, f_w, f_{ww}) = (c_0(w) - rw) f_w + \del {f_w^2 \over f_{ww}}.
$$
Note that $F$ is independent of $f$ and is non-increasing with respect to $f_{ww}$ and thereby satisfies the monotonicity condition in Crandall et al.\ (1992, (0.1)).  We have $\phio(\al m, m) = 1 = \phil(\al m, m)$, $\phio \left( \ws^{(0)}, m \right) = 0 \le \phil \left( \ws^{(0)}, m \right)$ because $\ws^{(0)} \le \ws^{(1)}$, $F \left(w, \phio, \phio_w, \phio_{ww} \right) = 0$, and
$$
F \left(w, \phil, \phil_w, \phil_{ww} \right) = - (c_1(w) - c_0(w)) \phil_w \ge 0,
$$
because $\phil_w \le 0$.  Thus, by comparison of viscosity solutions (Crandall et al., 1992, Theorem 3.3), it follows that $\phio \le \phil$ on $\D^{(0)}$.  \qed

\rem{4.3} {The comparison result in Lemma 4.2 also follows from a probabilistic argument.  Indeed, for any fixed admissible investment strategy, the probability of drawdown increases with the payout rate; denote this inequality by $h^{(0)}(w, m; \pi) \le h^{(1)}(w, m; \pi)$.  Then, by taking the infimum of $h^{(0)}(w, m; \pi)$ over admissible investment strategies, we have $\phio(w, m) \le h^{(1)}(w, m; \pi)$. Finally, by taking the infimum of $h^{(1)}(w, m; \pi)$ over admissible investment strategies, we have $\phio(w, m) \le \phil(w, m)$.  \qed}

\prop{4.3} {Suppose $\ws = \infty$, and suppose there exist $L > 0$ and $w_0$ such that $c(w) - rw > L$ for all $w > w_0$.  Then, the minimum probability of drawdown is identically $1$.}

\pf First, suppose that $c(w) - rw$ is eventually bounded.  Specifically, suppose there exist $L > 0$ and $L' > L$ such that for $\eps \in (0, L)$, there exists $w' > 0$ such that 
$$
L - \eps < c(w) - rw < L' + \eps,
$$
for all $w > w'$.  Then, for $\al m > w'$, it follows that
$$
- \int_{\al m}^y {\del \, du \over L - \eps} < - \int_{\al m}^y {\del \, du \over c(u) - ru} < - \int_{\al m}^y {\del \, du \over L' + \eps},
$$
or equivalently,
$$
- {\del \over L - \eps} \, (y - \al m) < - \int_{\al m}^y {\del \, du \over c(u) - ru} < - {\del \over L' + \eps} \, (y - \al m).
$$
It follows that
$$
\int_{\al m}^{m} e^{- {\del \over L - \eps} (y - \al m) } \, dy < \int_{\al m}^{m} \exp \left( - \int_{\al m}^y {\del \, du \over c(u) - ru}  \right) \, dy < \int_{\al m}^{m} e^{- {\del \over L' + \eps} (y - \al m) } \, dy,
$$
or equivalently,
$$
{L - \eps \over \del} \left[ 1 - e^{-{\del \over L - \eps} \, m(1 - \al)} \right] < g(m, m) < {L' + \eps \over \del} \left[ 1 - e^{-{\del \over L' + \eps} \, m(1 - \al)} \right]
$$
Thus, because $\eps > 0$ is arbitrary, $\lim_{m \to \infty} g(m, m) \in \left[{L \over \del}, {L' \over \del} \right]$.  

Next, we show that the minimum probability of drawdown is identically $1$ in this case.  To that end, note that, for $\al y > w'$, we have
$$
{L - \eps \over \del} \left[1 - e^{- {\del \over L - \eps} \, y (1 - \al)} \right] < g(y, y) < {L' + \eps \over \del} \left[1 - e^{- {\del \over L' + \eps} \, y (1 - \al)} \right],
$$
and
$$
{\del \over L' + \eps} < {\del \over c(\al y) - r \al y} < {\del \over L - \eps}.
$$
Thus, from (3.5),
$$
 {\al \del \over L' + \eps} \, {1 \over 1 - e^{- {\del \over L' + \eps} \, y (1 - \al)}} - {\al \del \eps \over L - \eps} < f(y) < {\al \del \over L - \eps} \, {1 \over 1 - e^{- {\del \over L - \eps} \, y (1 - \al)}} - {\al \del \eps \over L' + \eps} ,
$$
for $\al y > \al m > w'$, which implies that
$$
\left( {e^{{\del \over L - \eps} \, m(1 - \al)} - 1 \over e^{{\del \over L - \eps} \, N(1 - \al)} - 1} \right)^{{\al \over 1 - \al}} \, e^{-{\al \del \eps \over L' + \eps} \, (N - m)} < e^{- \int_m^N f(y) \, dy} < \left( {e^{{\del \over L' + \eps} \, m(1 - \al)} - 1 \over e^{{\del \over L' + \eps} \, N(1 - \al)} - 1} \right)^{{\al \over 1 - \al}} \, e^{-{\al \del \eps \over L - \eps} \, (N - m)},
$$
for $N > m$.  By letting $N \to \infty$, we see that $k(m)$ in (3.7) is zero  when $\al m > w'$.  Furthermore, if $\al m \le w_0$, then
$$
k(m) = e^{-\int_m^\infty f(y) \, dy} = e^{-\int_m^{w_0/\al} f(y) \, dy} \, e^{-\int_{w_0/\al}^\infty f(y) \, dy} = 0,
$$
so the minimum probability of drawdown $\phi$ is identically $1$.

Now, suppose the more general condition on $c(w) - rw$ as stated in the proposition.  Choose $L_1 > L$, and define $c_1$ by $c_1(w) = \min(c(w), \, L_1 + rw)$.  Then, from the above argument, $\phi_1$, the minimum probability of drawdown when the payout function equals $c_1$ is identically $1$.  Lemma 4.2 implies that $\phi \ge \phi_1$, in which $\phi$ is the minimum probability of drawdown under the original payout function $c$; thus, $\phi$ is identically $1$.  \qed
 
\rem{4.4} {In some sense, Proposition 4.3 is an analog of Proposition 4.1 in the case for which $\ws = \infty$.  Indeed, one important case {\it not} covered by Proposition 4.1 is the one for which $c(w)$ is tangent to $rw$ as $w \to \ws^-$.  Similarly, one important case {\it not} covered by Proposition 4.3 is the one for which $\lim_{w \to \infty} c(w) - rw = 0$, which requires that $c(w)$ approach $rw$ asymptotically as $w \to \infty$. \qed}



\bigskip

\centerline{\bf Acknowledgments}

\medskip

Research of the second author is supported in part by the National Science Foundation under grant DMS-0955463 and by the Susan M. Smith Professorship of Actuarial Mathematics. Research of the third author is supported in part by the Cecil J. and Ethel M. Nesbitt Professorship of Actuarial Mathematics.

\sect{References}

\noindent \hangindent 20 pt  Angoshtari, Bahman, Erhan Bayraktar, and Virginia R. Young (2015), Minimizing the expected lifetime spent in drawdown under proportional consumption, to appear in {\it Finance Research Letters}.

\smallskip \noindent \hangindent 20 pt Angoshtari, Bahman, Erhan Bayraktar, and Virginia R. Young (2016), Minimizing the probability of lifetime drawdown under constant consumption, working paper, University of Michigan.

\smallskip \noindent \hangindent 20 pt  B\"auerle, Nicole and Erhan Bayraktar (2014),  A note on applications of stochastic ordering to control problems in insurance and finance, {\it Stochastics}, 86(2): 330-340. 

\smallskip \noindent \hangindent 20 pt  Bayraktar, Erhan, and Virginia R. Young (2007), Correspondence between lifetime minimum wealth and utility of consumption, {\it Finance and Stochastics}, 11(2): 213-236.

\smallskip \noindent \hangindent 20 pt  Bayraktar, Erhan, and Virginia R. Young (2015), Optimally investing to reach a bequest goal, working paper, University of Michigan.

\smallskip \noindent \hangindent 20 pt Browne, Sid (1997), Survival and growth with a liability: optimal portfolio strategies in continuous time, {\it Mathematics of Operations Research}, 22 (2): 468-493.

\smallskip \noindent \hangindent 20 pt Browne, Sid (1999a), Beating a moving target: optimal portfolio strategies for outperforming a stochastic benchmark, {\it Finance and Stochastics}, 3 (3): 275-294.

\smallskip \noindent \hangindent 20 pt Browne, Sid (1999b), Reaching goals by a deadline: digital options and continuous-time active portfolio management, {\it Advances in Applied Probability}, 31 (2): 551-577.

\smallskip \noindent \hangindent 20 pt Crandall, Michael G., Hitoshi Ishii, and Pierre-Louis Lions (1992), User's guide to viscosity solution of second-order partial differential equations, {\it Bulletin of the American Mathematical Society}, 27 (1): 1-67.

\smallskip \noindent \hangindent 20 pt Dubins, Lester E. and Leonard J. Savage (1965, 1976), {\it How to Gamble if You Must: Inequalities for Stochastic Processes}, 1965 edition McGraw-Hill, New York. 1976 edition Dover, New York.

\smallskip \noindent \hangindent 20 pt Elie, Romuald and Nizar Touzi (2008), Optimal lifetime consumption and investment under a drawdown constraint, {\it Finance and Stochastics}, 12: 299-330.

\smallskip \noindent \hangindent 20 pt Karatzas, Ioannis (1997), Adaptive control of a diffusion to a goal, and a parabolic Monge-Ampere-type equation, {\it Asian Journal of Mathematics}, 1: 295-313.

\smallskip \noindent \hangindent 20 pt Karatzas, Ioannis and Steven E. Shreve (1991), {\it Brownian Motion and Stochastic Calculus}, New York: Springer-Verlag.

\smallskip \noindent \hangindent 20 pt Kulldorff, Martin (1993), Optimal control of favorable games with a time limit, {\it SIAM Journal on Control and Optimization}, 31 (1): 52-69.

\smallskip \noindent \hangindent 20 pt Orey, Steven, Victor C. Pestien, and William D. Sudderth (1987), Reaching zero rapidly, {\it SIAM Journal on Control and Optimization} 25 (5): 1253-1265.

\smallskip \noindent \hangindent 20 pt Pestien, Victor C. and William D. Sudderth (1985), Continuous-time red and black: how to control a diffusion to a goal, {\it Mathematics of Operations Research}, 10 (4): 599-611.

\smallskip \noindent \hangindent 20 pt Sudderth, William D. and Ananda Weerasinghe (1989), Controlling a process to a goal in finite time, {\it Mathematics of Operations Research}, 14 (3): 400-409.

\smallskip \noindent \hangindent 20 pt Young, Virginia R. (2004), Optimal investment strategy to minimize the probability of lifetime ruin, {\it North American Actuarial Journal}, 8 (4): 105-126.

\bye